\documentclass{aa}

\usepackage{natbib}
\usepackage{array}
\usepackage{graphicx,dblfloatfix}
\usepackage{epstopdf}
\usepackage{txfonts}

\begin{document}

   \title{Collision Avoidance in Next-generation Fiber Positioner Robotic System for Large Survey Spectrograph}

   \author{Laleh Makarem
          \inst{1,7}
          \and
          Jean-Paul Kneib\inst{2,3}
          \and
          Denis Gillet\inst{1}
          \and
          Hannes Bleuler\inst{4}
          \and
          Mohamed Bouri\inst{4}
          \and
          Laurent Jenni\inst{4}
          \and
          Francisco Prada\inst{5,6}
          \and
          Justo Sanchez\inst{5}
          }

   \institute{Coordination and Interaction Systems Group (REACT),
    Ecole Polytechnique F\'{e}d\'{e}rale de Lausanne (EPFL), Switzerland
         \and
             Laboratory of Astrophysics (LASTRO),
             Ecole Polytechnique F\'{e}d\'{e}rale de Lausanne (EPFL),
             Observatoire de Sauverny, Ch-1290 Versoix, Switzerland
         \and
            Aix Marseille Universit\'{e}, CNRS, LAM (Laboratoire d'Astrophysique de Marseille) UMR 7326, 13388, Marseille, France
         \and
            Laboratory of Robotic Systems (LSRO)
            Ecole Polytechnique F\'{e}d\'{e}rale de Lausanne (EPFL), Switzerland
        \and
         Instituto de Astrofisica de Andalucia (CSIC), Granada, E-18008, Spain
         \and
         Instituto de F\'{i}sica Te\'{o}rica, (UAM/CSIC), Universidad Aut\'{o}noma de Madrid, Cantoblanco, E-28049 Madrid, Spain
         \and
         \email{laleh.makarem@epfl.ch}    
             }


 
  \abstract
 {Some of the next generation massive spectroscopic survey projects, such as DESI and PFS, plan to use thousands of fiber positioner robots packed at a focal plane to quickly move in parallel the fiber-ends from the previous to the next target points. The most direct trajectories are prone to collision that could damage the robots and impact the survey operation. We thus present here a motion planning method based on a novel decentralized navigation function for collision-free coordination of fiber positioners. The navigation function takes into account the configuration of positioners as well as the actuator constraints. We provide details for the proof of convergence and  collision avoidance. Decentralization results in linear complexity for the motion planning as well as dependency of motion duration with respect to the number of positioners. Therefore the coordination method is scalable for large-scale spectrograph robots.  The short in-motion duration of positioner robots ($\sim$2.5 seconds using typical actuator constraints), will thus allow the time dedicated for observation to be maximized.}


   \maketitle
%
\section{Introduction}
After the discovery and confirmation of the accelerated expansion of the universe (\citealt{riess1998observational,perlmutter1999measurements}), one of the main challenges in cosmology is to discern the nature of the dark energy. In order to achieve this goal, different observational techniques have been proposed to tackle the geometry and evolution of the Universe. One of the key techniques is  the measurement of the Baryonic Acoustic Oscillations (BAO) in massive spectroscopic surveys.

The very first large-scale spectroscopic survey (\citealt{huchra1983survey}) revealed a cosmic web structure with filaments and voids, and soon after, further investigations  questioned the existence of a cosmological constant (\citealt{efstathiou1990cosmological}).
More recently, following the discovery of the imprint of the BAO in the Sloan Digital Sky Survey (SDSS; \citealt{eisenstein2005detection}), massive spectroscopic surveys have been developed to measure accurately the evolution of the distance-redshift relation using the BAO technique. In particular, 1) the WiggleZ redshift survey (\citealt{blake2011wigglez}) has completed a $\sim$250,000 redshift survey of star-forming galaxies (at $z<0.8$) at the 4m Anglo Australian Telescope (AAT), 2) the Baryonic Oscillation Spectroscopic Survey (BOSS; \citealt{anderson2012clustering}) will complete in 2014 a major redshift survey of 1.4 million galaxy redshift (at $z<0.7$) and 160,000 high-redshift Lyman-$\alpha$ quasars using the SDSS telescope, and 3) the extended-BOSS survey (2014-2020) will complete the first BAO survey over the redshift range $0.7<z<2.2$ using galaxies and quasars as well as the SDSS facility. 

To go beyond the throughput limits of current surveys, new technologies are being developed to fasten the future spectroscopic facilities. The way forward is not only to use larger aperture telescopes, but also to use a larger multiplexing. Over the past few years two major projects have been approved for construction. First, the Primary Focus Spectrograph (PFS) is a Japanese lead project that aims to develop a 2400-fiber spectrograph on the 8.2 m Subaru telescope. Second, the Dark Energy Spectroscopic Instrument (DESI)\footnote{http://desi.lbl.gov}, a DOE lead project,  aims to develop a 5000-fiber spectrograph on the Mayall 4m telescope. Other less advanced projects are also being prepared such as 4MOST and WEAVE.
\begin{figure*}[tb]
\label{FocalPlane}
  \centering
    \includegraphics[width=0.8\textwidth]{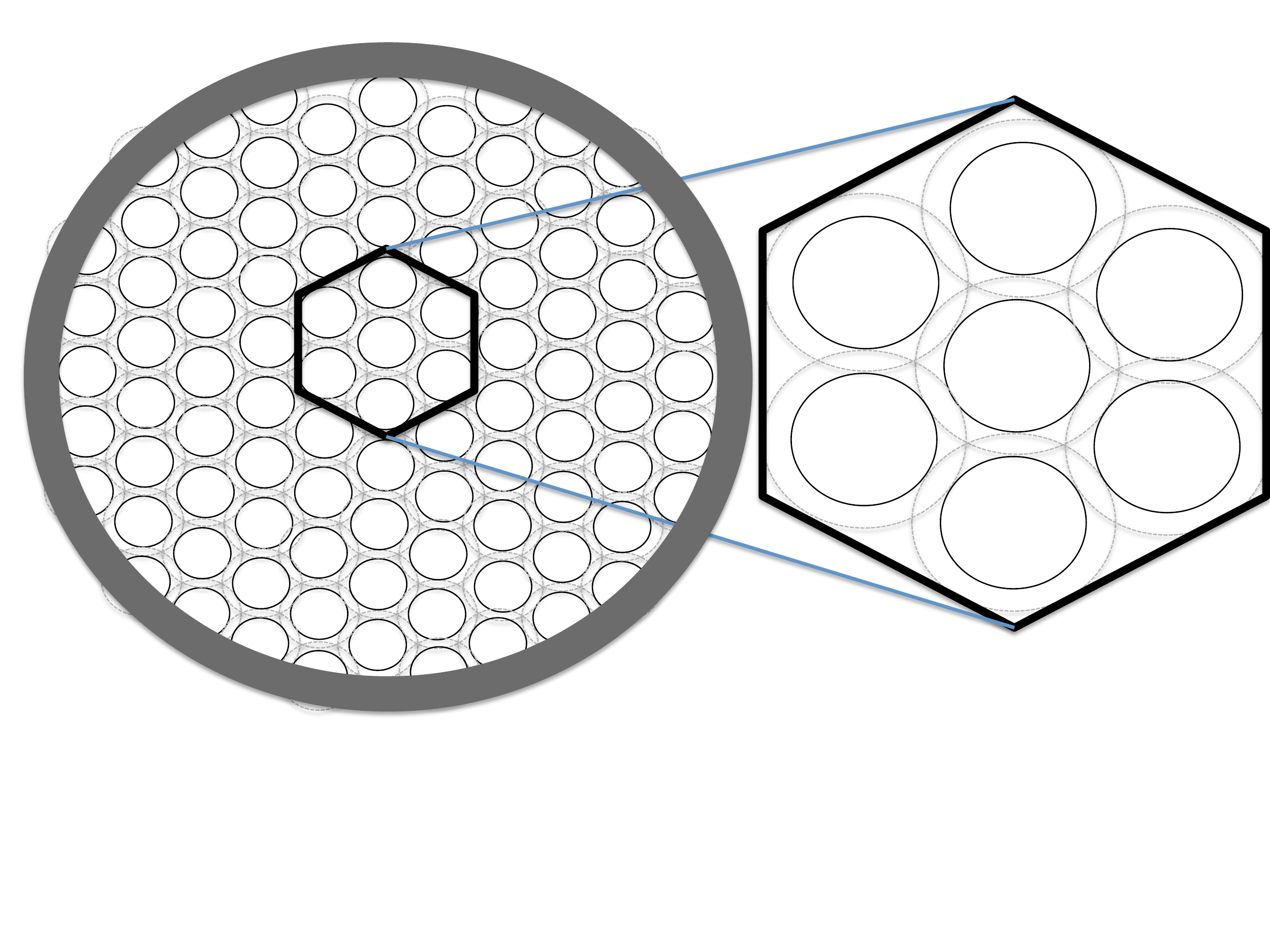}
  \caption{Configuration of robot positioner on a focal plane. Positioners shown as black circles, form a hexagonal array in order to cover a patrol discv(shown in a dashed gray circle) as versetile as possible.}
\end{figure*}

Spectrographs fed by massive fiber bundles are one of the most advanced and proven methods compared with multi-slit approach. Various technologies have been proposed for fiber positioning. For example, in the case of the SDSS spectrograph, fibers are placed manually into the holes drilled in an aluminum plate. This operation is done during the day time prior to observations. In the case of the AAT spectrograph a robot is placing a fiber after a fiber at the target point. This operation is done while another set of fiber is observing. In the case of the Chinese Large Sky Area Multi-Object Fibre Spectroscopic Telescope (LAMOST; \citealt{cui2010southern}), and the Japanese Fibre Multi-Object Spectrograph (FMOS; \citealt{kimura2003fibre}) robotic fiber positioners are placing in parallel the fibers at the target points just before the observations are conducted. The key advantage of using robotic positioners is that the fibers could be positioned simultaneously. So if the robotic system is fast, reconfiguration time could executed during the observation overheads (readout time of the detectors and slewing of the telescope).  

In next-generation fibre-fed spectrographs such as the one in DESI or PFS projects, small robots are responsible to position the fiber-ends. In order to improve the versatility of the system and ensure the maximum number of observed galaxies the robots share working spaces. For robotic systems using a two eccentric rotary joints ($\theta-\phi$ design), one of the challenges is to move a single fiber-end without having collision with the others.This work presents a new motion planning method for the positioner robots based on Decentralized Navigation Functions (DNF). The proposed trajectories guarantee collision-free path for all the fiber-ends. In addition, the motion planning is decentralized in order to be able to extend the solution for a large-scale positioner robots.

This paper is organized as follows. In section 2, a description of the focal plane is given. In section 3, we give the problem formulation for a collision-free trajectory planner. The solution to this problem, using DNF, is given in section 4.  Proof of convergence and constraints on parameter tuning is explained in section 5. The simulation results corresponding to the proposed approach are presented and discussed in section 6. Section 7 briefly explores some avenues for future research and concludes.

\section{Description of focal plane}

We describe a standard design for the focal plane, which can be extended to a number of future fibre-fed multi-object spectrograph instruments (and particularly designs explored for DESI and PFS).  The main concept is a collection of identical positioners distributed over an hexagonal array (See Fig. \ref{FocalPlane}). 
Each positioner robot is therefore assigned to a single target or if no target is accessible will observe the sky. 
The positioner robots could cover the entire focal plane and each move a fibre head toward a desired location within the patrol disc of positioner. Because we require that any point of the focal plane is accessible by a fiber, there will be regions of the focal plane where the workspaces of adjacent positioners will overlap. In these overlap zones  there is a risk of actuator collision. Target assignment and collision avoidance strategies will therefore be among the challenges in the design of such a massively parallel fibre-fed spectrograph. 

\subsection{Target assignment}

Several strategies can be developed in order to assign targets -galaxies, quasars or stars- to the end effectors -fiber-ends- of the positioner robots. The simplest approach is to select randomly for each positioner any of the targets lying within the corresponding working space (patrol disc). To achieve an optimal solution which means that the maximum number of targets is observed during a certain period of time, the drain algorithm was introduced (\citealt{morales2012fibre}). The method ensures observation of  maximum number of targets in the first few sets of observations. Using both randomly distributed targets and mock galaxy catalogues,  the authors showed that the number of observed galaxies could be increased by 2 percents in the first set of observations.

In this paper we assume that target assignment has been effectively done by one of the mentioned algorithms. Thus, the focus of the work presented here is on the coordination of positioners in motion to avoid collision. We assume that the target of each positioner is fixed (not a quickly moving target) and known to the positioner robot.

\section{Problem formulation}
\begin{figure}[td]
\label{ThetaPhi}
  \centering
    \includegraphics[width=0.5\textwidth]{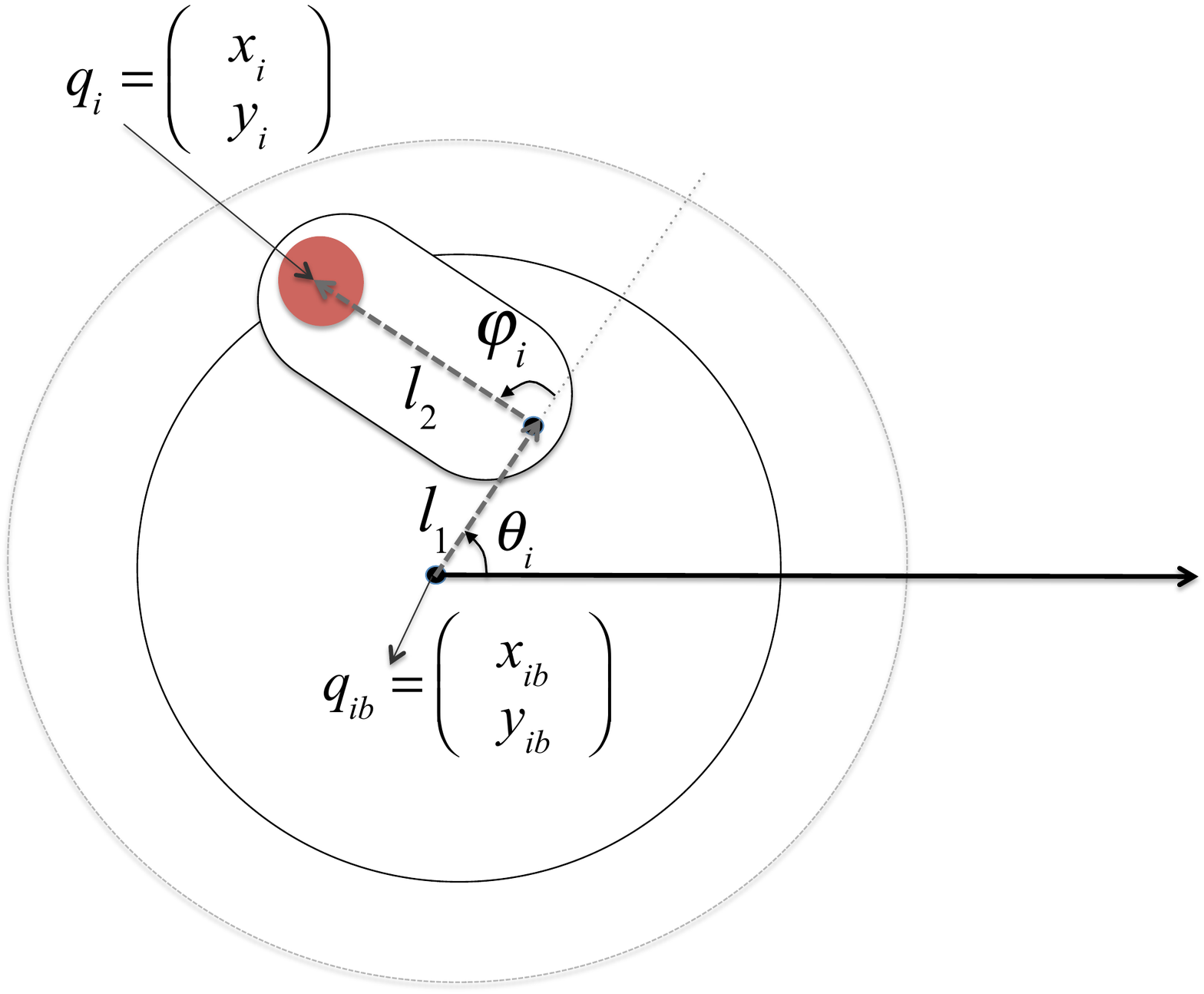}
  \caption{A positioner robot with two degrees of freedom. The main disk (black circle) rotates along its central axis. Its angular position is shown as $\theta_{i}$. The arm with the length of $l_2$ rotates around an eccentric axis (with the distance of $l_1$ from the center) fixed on the main disk and its angular position is shown as $\phi_{i}$. $q_{ib}$ is the position of the robot base fixed to the telescope structure in a global frame attached to the focal plane. $q_i$ is the position of the fiber attached to the robot $i$ in that global frame.}
\end{figure}

We consider a system composed of $N$ positioner robots. The goal of each robot is to put its end effector (fiber-end) on an assigned target point. Each positioner is a planar robot with two degrees of freedom, each moving by a rotational motor (Fig. \ref{ThetaPhi}).

Each positioner robot covers the patrol area (workspace) through two correlated rotations: $\theta$ rotates around the actuator's main axis, while $\phi$ moves the fiber arm tip from the peripheral circle to the central axis (Fig. \ref{ThetaPhi}). The optical fiber is attached to a ferrule at the arm tip, and passes through the center of the actuator and exits by the rear side of the robot.

The forward kinematics of each positioner robot can be described as:
\begin{equation}
\label{DynamicModel}
\left( {\begin{array}{*{20}c}
   {x_i }  \\
   {y_i }  \\
\end{array}} \right) = \left( {\begin{array}{*{20}c}
   {x_{ib} }  \\
   {y_{ib} }  \\
\end{array}} \right) + \left( {\begin{array}{*{20}c}
   {cos\theta_{i} } & {cos(\theta_{i}  + \phi_{i} )}  \\
   {sin\theta_{i} } & {sin(\theta_{i}  + \phi_{i} )}  \\
\end{array}} \right)\left( {\begin{array}{*{20}c}
   {l_1 }  \\
   {l_2 }  \\
\end{array}} \right)
\end{equation}

Where the end-effector position of positioner $i$ is $q_i = (x_i,y_i)$  in a global frame attached to the focal plane. $l_1$ and $l_2$ are first and second rotation links respectively (Fig. \ref{ThetaPhi}).  $\theta$ and $\phi$ are angular positions of the two joints of the positioner $i$. Each positioner is controllable by its angular velocity, meaning the speed of each of the two motors.

The main challenge is to coordinate the robots in motion to reach a predefined target point while avoiding collisions. The proposed approach should be expandable to a more large-scale problems like the one with 5000 fiber-positioners in the DESI or the 24000 fiber-positioners of PFS. A centralized solution for such a problem would be practically infeasible and costly due to the presence of numerous positioners and constraints (\citealt{tanner2005towards}). Therefore, among all the methods found in the literature for coordinating agents, decentralized and reactive control approaches seem promising.

\section{Decentralized Navigation Function}

Inspired by the emergent behaviors among swarms (insects, birds, fishes), methods based on local reactive control have received great interest (\citealt{ge2012decentralized}). Therefore, artificial potential fields are often exploited for the coordination of autonomous agents. The main drawback of most potential field approaches is that convergence to the target is not guaranteed due to the presence of spurious local minima. In order to solve this problem and present a complete exact solution for the coordination problem, navigation functions have been introduced (\citealt{dimarogonas2005decentralized}).
\begin{figure}[t]
\label{FiberGoal}
  \centering
    \includegraphics[width=0.5\textwidth]{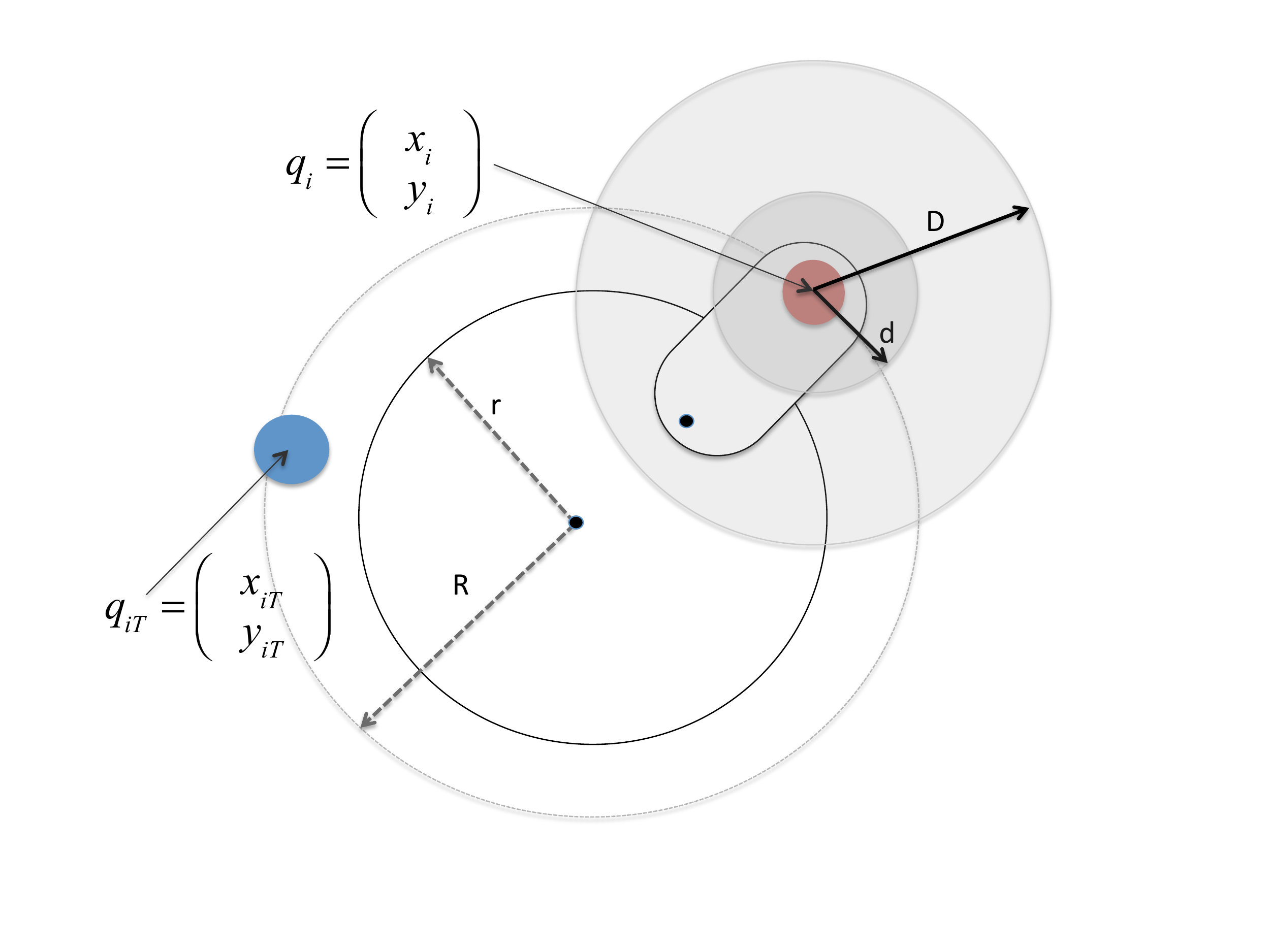}
  \caption{The target point $q_{iT}$ is the destination of the robot end-effector.The collision detection envelope with radius of D, is the area where collision avoidance term in the navigation function is activated.The collision avoidance term in the navigation function ensures that the two positioner end-effectors will keep a distance larger than d.}
\end{figure}
\begin{figure}[t]
\label{TwoRobots}
  \centering
    \includegraphics[width=0.5\textwidth]{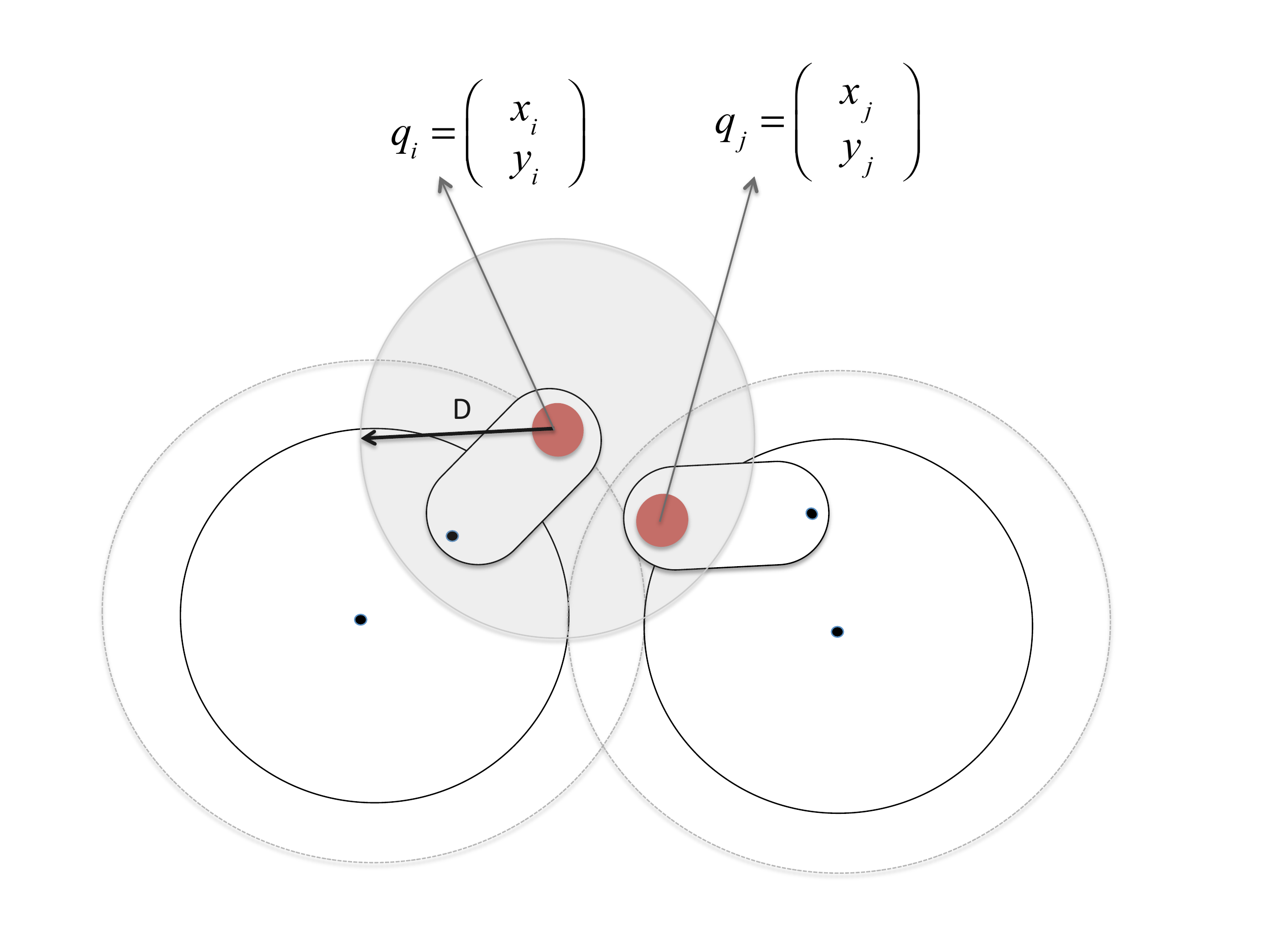}
  \caption{A configuration in which there is a risk of collision between the end-effectors of two robots. In this configuration the collision avoidance term in the navigation functions of the two robots are active which means they take values more than zero.}
\end{figure}
Navigation functions have been used in various robotic and control applications (\citealt{makarem2011decentralized, makarem2012fluent, de2006formation}). In these applications the actuation torque or other inputs (e.g., the acceleration, the velocity) is derived from some potential function that encodes relevant information about the environment and the objective. In the framework of the problem presented in this paper, the use of navigation functions in a decentralized scheme is promising, as it can be implemented in real-time and it also shows good flexibility with regard to adding new robots and changing the problem constraints.

A navigation function is practically a smooth mapping which should be analytic in the workspace of every positioner robot and its gradient would be attractive to its destination and repulsive from other robots. So, an appropriate navigation function could be combined with a proper control law in order to obtain a trajectory for every motor of robot leading to the destination and avoiding collisions.

We propose a navigation function (\ref{DNF}) which is composed of two components. The first part, the attractive term, is the squared distance of the end-effector of positioner robot $i$ from its target point. This term of navigation function attains small values as the positioner brings the fiber closer to its target point(Fig. \ref{FiberGoal}). The second part, the repulsive term, aims at avoiding collisions between positioner $i$ and the six other positioners located in its vicinity. This term is activated when the two positioner robots are closer than a distance $D$, otherwise this term is zero. $D$ defines the radius of a collision avoidance envelope and $d<D$ defines the radius of the safety region. The closer the two robots get, this repulsive term attains higher values. Moving toward the minimum point of this navigation function will guarantee the minimum distance of $d $ between the positioners (Fig. \ref{TwoRobots}). 

\begin{equation}
\label{DNF}
\psi _i  = \lambda_1 \left\| {q_i  - q_{iT} } \right\|^2 + \lambda_2 \sum\limits_{j \ne i} {min(0,\frac{{\left\| {q_i  - q_j } \right\|^2  - D^2 }}{{\left\| {q_i  - q_j } \right\|^2  - d^2 }})} 
\end{equation}

According to the navigation function presented in (\ref{DNF}) and the forward kinematics defined in (\ref{DynamicModel}), the following control law is proposed:
\begin{equation}
\label{Control}
u_i  =  - \nabla _{\theta _i  \phi_i} \psi _i (q_i )
\end{equation}
In every step, the robot will move the fiber according to a gradient descent method. It is worth mentioning that, the navigation function is directly a function of the end-effector positions. In order to obtain the angular velocities of each of two motors, we calculate the gradient of the navigation function with respect to the angular positions of links using the chain derivatives and forward kinematics in (\ref{DynamicModel}).\\
\begin{equation}
\label{ChainDer}
\left( {\begin{array}{*{20}c}
   {\omega _{i1} }  \\
   {\omega _{i2} }  \\
\end{array}} \right) = u_i  =  - \left( {\begin{array}{*{20}c}
   {\frac{{\partial \psi _i }}{{\partial x_i }}\frac{{\partial x_i }}{{\partial \theta_{i} }} + \frac{{\partial \psi _i }}{{\partial y_i }}\frac{{\partial y_i }}{{\partial \theta_{i} }}}  \\
   {\frac{{\partial \psi _i }}{{\partial x_i }}\frac{{\partial x_i }}{{\partial \phi_{i} }} + \frac{{\partial \psi _i }}{{\partial y_i }}\frac{{\partial y_i }}{{\partial \phi_{i} }}}  \\
\end{array}} \right)
\end{equation}
$\omega_{i1}$ and $\omega_{i2}$ are the angular velocity of the first and second motor of positioner robot $i$ respectively.

\subsection{Algorithm complexity}
\begin{table*}[t]
\center
\label{Table1}
\caption{Motion planning algorithm for all positioner robots. $T_0$ and $T_f$ are the beginning and end times of the algorithm respectively. $dt$ is the time step and M is the number of time steps i.e. $M=\frac{T_f - T_0}{dt}$}
\begin{tabular}{ll}
\hline
\hline
\multicolumn{2}{c}{{\bf Trajectory planning algorithm}}\\
\hline
{\bf Inputs:} &Initial end-effector position of all the positioners: $Q_{init} = [q_1,q_2,\ldots,q_N]$ \\
&and target position of all fibers assigned to each positioner: $Q_{goal} =  [q_{1T},q_{2T},\ldots,q_{NT}]$\\ 
{\bf outputs:} & A sequence of motor peed values for each positioner\\
&$\Omega_1 = [\omega_1(1),\ldots,\omega_1(M)]$\\
&$\Omega_2 = [\omega_2(1),\ldots,\omega_2(M)]$\\
&$\vdots$\\
&$\Omega_N = [\omega_N(1),\ldots,\omega_N(M)]$\\
\hline
\multicolumn{2}{l}{m = 0}\\
\multicolumn{2}{l}{{\bf repeat until $\nabla \psi =0$}}\\
\multicolumn{2}{l}{$\qquad${\bf for each positioner(i=1:N)}}\\
\multicolumn{2}{l}{$\qquad \qquad \omega_i(m) =  - \nabla \psi (q_i(m), q_{iT}, Q(m))$ See: Table. 2}\\
\multicolumn{2}{l}{$\qquad \qquad q_i(m + 1) = q_i(m + 1) + dt . \omega_i(m)$ }\\
\multicolumn{2}{l}{$\qquad \qquad$m = m + 1}\\
\multicolumn{2}{l}{$\qquad \qquad\nabla \psi =$ max$_i(\nabla \psi (q_i(m), q_{iT}, Q(m)))$}\\
\multicolumn{2}{l}{$\qquad${\bf end for}}\\
\multicolumn{2}{l}{{\bf end repeat}}\\
\hline
\hline
\end{tabular}
\end{table*} 
\begin{table*}[t]
\center
\label{Table2}
\caption{Motion planning algorithm calls this module that calculates gradient of the DNF. This function gets the current and target position of the robot as well as current position of its adjacent positioners. The output of the function is the gradient of the DNF.}
\begin{tabular}{ll}
\hline
\hline
\multicolumn{2}{c}{{\bf Gradient of the DNF for the positioner $i$}}\\
\hline
{\bf Inputs:} & Current position of the positioner $q_i$,\\
&target position of the positioner $q_{iT}$,\\
&and current position of the neighbor positioners $Q_{neighbor} \in Q $\\
{\bf outputs:} & The gradient of the navigation function for positioner $i$ \\
&which is a vector of a two elements $G = \left( {\begin{array}{*{20}c}{G_1 }  \\ {G_2 }  \\ \end{array}} \right) $\\
\hline
\multicolumn{2}{l}{$G = 0$}\\
\multicolumn{2}{l}{$G = G + 2\lambda_1(q_i-q_{iT})$}\\
\multicolumn{2}{l}{{\bf for each neighbor positioner(j=1:6)}}\\
\multicolumn{2}{l}{$ \qquad G = G + 2 \lambda_2 \frac{(R^2  - r^2 )(q_i  - q_j )}{(\left\| {q_i  - q_j } \right\|^2  - r^2 )^2 }$}\\
\multicolumn{2}{l}{{\bf end for}}\\
\hline
\hline
\end{tabular}
\end{table*} 

Table 1 describes the motion planning algorithm. In each time step $dt$, each positioner robot computes the speed of its two motors knowing its current position and the target point as well as the position of adjacent robots. In each time step, each positioner calls the module that computes the gradient of decentralized navigation function(Table 2). So, the algorithm runs in a for loop as many times as the number of positioner robots. On the other hand, the inner loop that calculates the gradient of the DNF runs constant times (number of adjacent positioners = 6). Therefore the complexity of the algorithm will remain O(N), where N is the number of positioner robots. Considering the fact that all the robots' bases are fixed to the focal plane, collisions can occur locally and chances of collision is only with the certain adjacent robots. Decentralizing motion planning and trajectory generation takes advantage of limited number of adjacent robots and the locality of collisions and significantly reduces the complexity of the algorithm to a linear order. Low complexity of the algorithm guarantees its ease of use in mid-scale and large-scale robotic telescopes where there are thousands, or tens of thousands of positioners respectively. 
 
\section{Collision-Free Motion Toward the Equilibrium}
\begin{figure*}[b]
\label{SimSnapShot}
  \centering
    \includegraphics[width=1\textwidth]{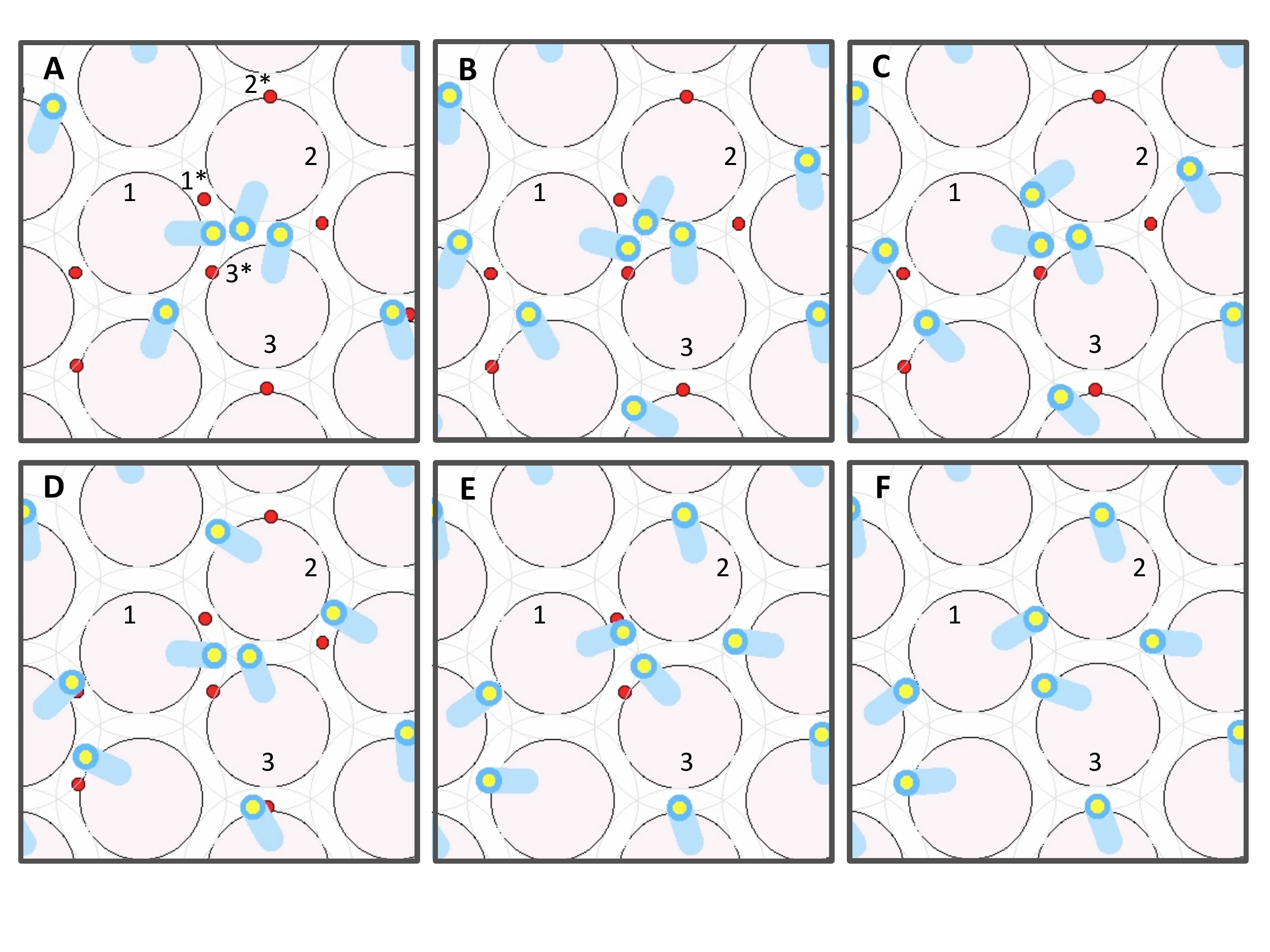}
  \caption{The six boxes (A to F) show six snap shots of the simulation. 1*, 2* and 3* are respectively the target positions for positioner 1,2 and 3. This three positioners are engaged in a local conflict in which positioner 1 needs to make space for positioner 2 to pass. positioner 2 can not make room for positioner 1 because positioner 3 is blocking the way. Positioner 3 needs to pass both positioners to reach its target point. The small maneuver from positioner 1 that comes directly from DNF moves this positioner farther from its target point but it makes room for the positioner 2 to pass safely. When positioner 2 clears the way, positioner 1 starts moving toward its target point and this gives a safe way to the positioner 3.}
\end{figure*}

The following theorem provides conditions under which the DNF (\ref{DNF}) in combination with the control law (\ref{Control}) ensures convergence of all robots to their target points. Convergence of all robots means that using this method, practically no blocking will occur even if some complex maneuvers are needed in case of intricate initial configurations of the robots (See simulation example in Fig.~\ref{SimSnapShot}). 

{\bf Theorem 1.} {\it If the following inequality is satisfied:
\begin{equation}
\frac{{\lambda _1 }}{{\lambda _2 }} < \frac{1}{R}\frac{D}{{D^2  - d^2 }}
\end{equation}

then the function (\ref{DNF}) is a Morse function and there is no local minimum except the equilibrium in the target point.}

{\bf Proof.} When $\nabla _{\theta _i } \psi (q_i ) = 0$ 

It is either in the presence of no other positioner in the vicinity of positioner $i$, where $\left\|q_i- q_j  \right\| \ge D$ or there is a risk of collision ${\left\| {q_i  - q_j } \right\| < D}$ with at least one other positioner. The gradient of the navigation function for both cases is: 

\begin{equation}
\nabla \psi (q_i ) = \left\{ {\begin{array}{*{20}c}
   {2\lambda _1 (q_i  - q_{iT} ) = 0} & {\left\| {q_i  - q_j } \right\| \ge D}  \\
   {\lambda _1 (q_i  - q_{iT} ) + 2\lambda _2 \sum\limits_{j = 1}^6 {\frac{{(D^2  - d^2 )(q_i  - q_j )}}{{(\left\| {q_i  - q_j } \right\|^2  - d^2 )^2 }}} } & {\left\| {q_i  - q_j } \right\| < D}  \\
\end{array}} \right.
\end{equation}

In case when there is no other positioner near the positioner $i$, $\nabla _{\theta _i } \psi (q_i ) = 0$ means $2\lambda _1 (q_i  - q_{iT} ) = 0$ which directly indicates that the positioner robot is in its target point. Otherwise, there is at least one other positioner in the collision avoidance envelope (D):

\begin{equation}
\nabla \psi (q_i ) = \lambda _1 (q_i  - q_{id} ) + 2\lambda _2 \sum\limits_{j = 1}^6 {\frac{{(D^2  - d^2 )(q_i  - q_j )}}{{(\left\| {q_i  - q_j } \right\|^2  - d^2 )^2 }}} 
\end{equation}

The first term in the gradient of the potential field always satisfies the following inequality:
\begin{equation}
\lambda _1 \left\| {q_i  - q_{id} } \right\| <2 \lambda _1 R
\end{equation}

In addition, the second term in the gradient of the potential field is:
\begin{equation}
\begin{array}{l}
\frac{{(D^2  - d^2 )\left\| {(q_i  - q_j )} \right\|}}{{(\left\| {q_i  - q_j } \right\|^2  - d^2 )^2 }} > \frac{D}{{D^2  - d^2 }}  \\ 
\to \lambda _2 \sum\limits_{j = 1}^6 {\frac{{(D^2  - d^2 )\left\| {(q_i  - q_j )} \right\|}}{{(\left\| {q_i  - q_j } \right\|^2  - d^2 )^2 }}}  > \lambda _2 \frac{2D}{{D^2  - d^2 }} \\ 
 \end{array}
\end{equation}

So, if $\lambda _2 \frac{2D}{{D^2  - d^2 }} > 2\lambda _1R$ then there is no point where the gradient of the potential field could be zero except at the target point. This guarantees that there will be no blocking (also called deadlock) in the method. 

\subsection{Parameter tuning}

There are two weighting parameters that can be tuned in DNF (\ref{DNF}); $\lambda_1$ and $\lambda_2$.  Theorem 1 gives a condition for tuning this parameters which guarantees convergence of all positioners to their target point. In order to ensure the success of the motion planning algorithm, the maximum velocity generated should not exceed the maximum velocity feasible for motors.  Lower values of $\lambda_1$ and $\lambda_2$ will result in lower values for velocity of motors and increase the convergence time. For an application in which fast convergence is desirable, the two parameters will be tuned to the highest values that still keep maximum generated velocity in the feasible range. We soft-tuned these two parameters through simulations.  

\section{Simulation Results}
\begin{figure*}[t]
\label{Profile}
  \centering
   \includegraphics[width=1\textwidth]{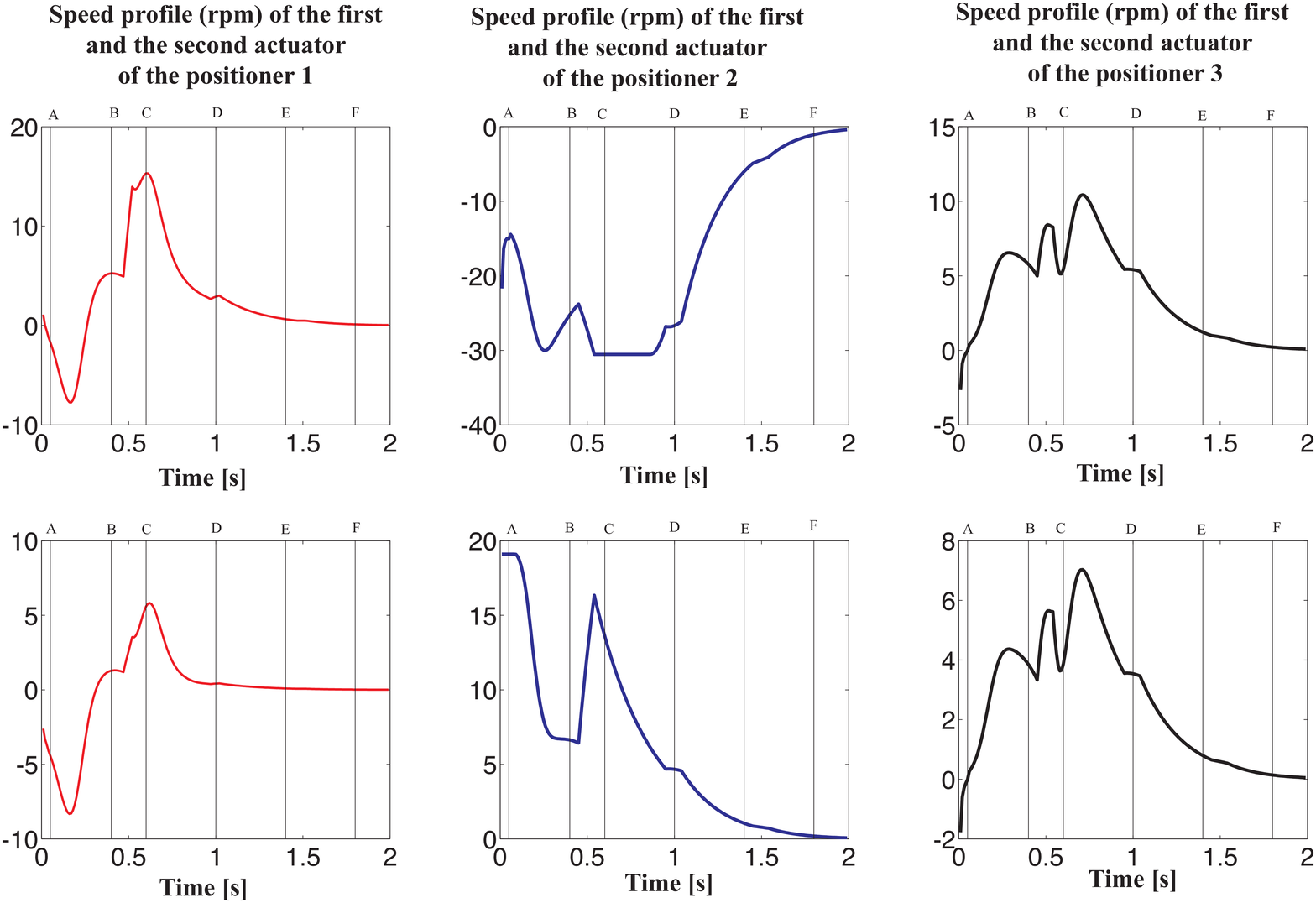}
  \caption{Velocity profiles that correspond to the pairs of actuators for positioners 1,2 and 3 in Fig. 5. Columns show the velocity profiles for each positioner. The first and second profiles of each column correspond to the first and second actuator of each positioner respectively. Vertical lines indicate the moment at which the snapshots of Fig. 5 were taken.}
\end{figure*}

In order to study the performance of the motion planning algorithm, we conducted various sets of simulations  for different number of positioners all in hexagonal configuration patterns. The size of positioners and the share volume between positioners are selected in a way to be compatible with next-generation spectrograph robots such as the one in Desi (Table 3).  
\begin{table}[b]
\center
\label{SimSpec}
\caption{Specifications of positioner robots (see Fig. 3 and Fig. 4)}
\begin{tabular}{>{\centering\arraybackslash}m{3 cm}  >{\centering\arraybackslash}m{3 cm}}
\hline
\hline
Parameter & Value\\
\hline
R & 7 mm\\
r & 5 mm\\
ferule size & 1 mm\\
arm diameter & 2 mm\\
arm length  & 3mm\\
D & 4 mm\\
d & 2 mm\\
\hline
\hline
\end{tabular}
\end{table} 

In all cases, initial angular position of the two motors of each positioner robot and their target points (the galaxies that positioners will observe) are randomly generated.  The reason for selecting different numbers of positioners is that we can verify our expectation on complexity of the algorithm. In addition, we can extrapolate the simulation time and motion duration for other number of positioners for larger scales of robotic spectrographs. For each set we repeated the simulations 100 times. Table.~\ref{SimParam} shows simulation parameters. The two weighting parameters in DNF (\ref{DNF}) should satisfy the condition in Theorem 1. According to positioners' specifications in Table.~\ref{SimSpec} $\lambda_1$ and $\lambda_2$ should fit the inequality of $\frac{{\lambda _1 }}{{\lambda _2 }} < \frac{1}{R}\frac{D}{{D^2  - d^2 }}$. This means $\lambda_1 < 72  \lambda_2$. 

\begin{figure}[t]
\label{Results1}
  \centering
    \includegraphics[width=0.5\textwidth]{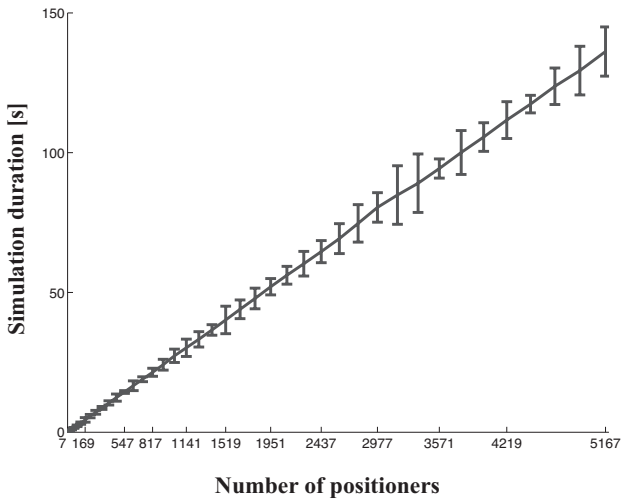}
\caption{Mean value simulation duration for sets with different number of positioners.  Repeating sets of simulation, the simulation durations do not significantly vary. The lengths of the error bars are therefore chose 10 times of the standard deviation at each point.}
\end{figure}

\begin{figure*}[t]
\label{Results2}
  \centering
    \includegraphics[width=1\textwidth]{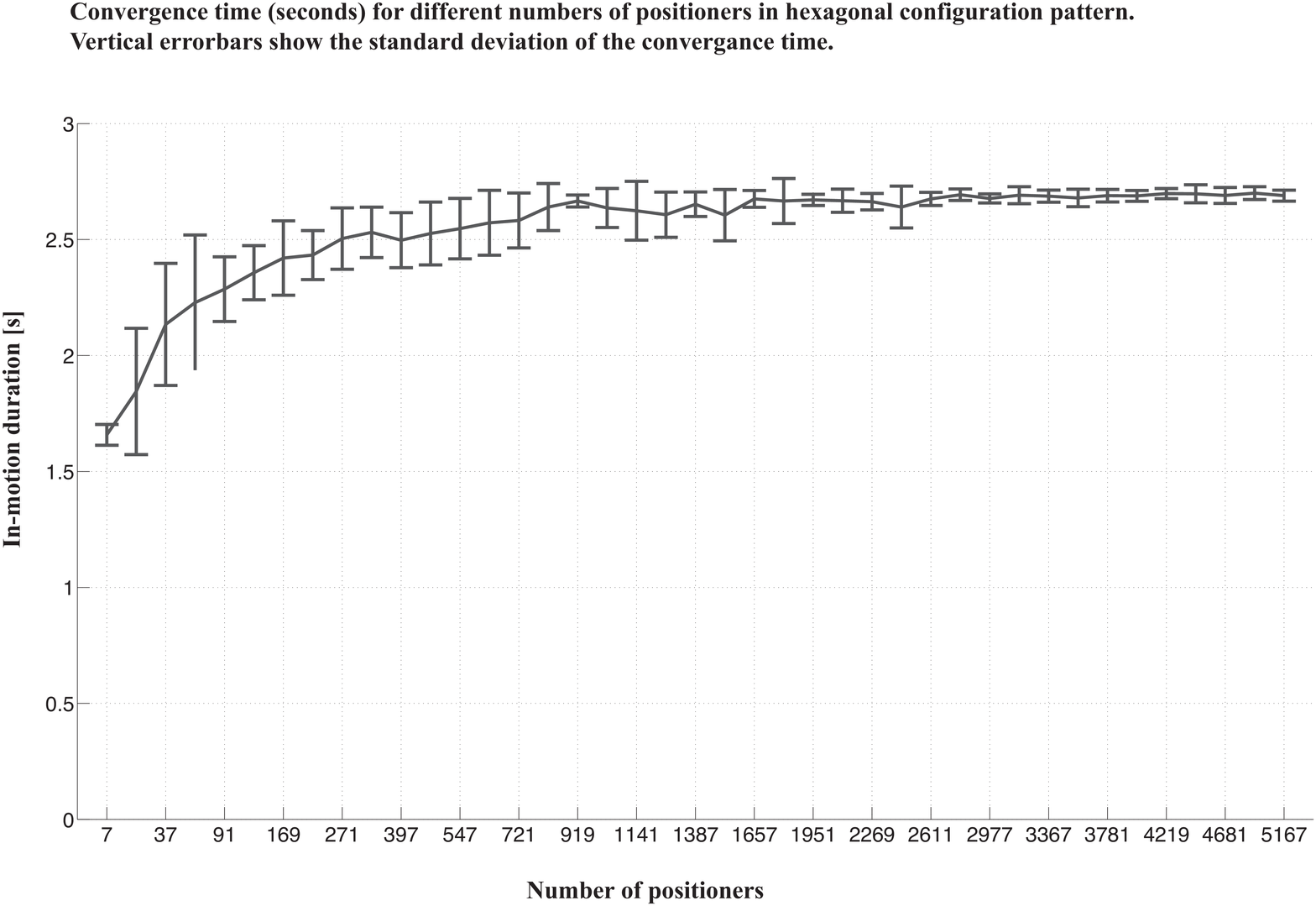}
\caption{Mean value and standard deviation of convergence time (seconds) for simulation sets with different numbers of positioners.Note that the x-axis is not linear with the number of positioners.}
\end{figure*}

As expected, we observe no collision during all sets of simulations (4100 sets). Fig.~\ref{Results1} and Fig.~\ref{Results2} show the simulation time and in-motion duration of robots respectively. In-motion duration is the convergence time needed for all the positioner robots to arrive in their target points considering constraints on actuator velocities in Table.~\ref{SimParam}.  Regarding values for simulation time, the processor for all simulations is a 3.33 GHz 6-Core Intel Xeon. 

\begin{table}[t]
\center
\label{SimParam}
\caption{Simulation parameters used for all sets of simulations}
\begin{tabular}{>{\centering\arraybackslash}m{3 cm}  >{\centering\arraybackslash}m{3 cm}}
\hline
\hline
Parameter & Value\\
\hline
Maximum speed of the first acuator & 30 rpm\\
Maximum speed of the second acuator & 20 rpm\\
$\lambda_1$ & 1 \\
$\lambda_2$ & 0.05\\ 
dt (time steps) & 10 ms\\
Convergence distance & 100 $\mu$m\\
\hline
\hline
\end{tabular}
\end{table} 
As expected , simulation time increases in a near linear manner with the number of positioners. This enables us to use this method for immediate motion planning for thousands of positioners. From results, the amount of time needed to run the simulation for 5000 positioners is less than 3 minutes (140 seconds) on today's typical computers. This amount of time is very reasonable as it is smaller than the typical exposure time foreseen for the DESI and PFS experiment, thus allowing to have an interactive planning of the observation. This would allow to modify in real time the target list due to many reasons, for example meteorological disturbances or even the last minute discovered transient targets such as SuperNovae or Gamma Ray Burst.

The main advantage of the method that derives from the idea of decentralization is that the motion duration does not change with the number of positioner robots. Decentralization benefits from the configuration of positioner robots and the fact that collisions could happen locally and they do not affect the motion of non-neighbor positioners. With realistic actuator constraints, as used in simulation sets, we can expect to accomplished the first run of motion for positioners of a mid-scale robot positioners in less than  2.5 seconds. Such small duration of time for coordination of all positioners will allow to maximize the duration of observation and survey efficiency.  

So far in all sets of simulations, we study the performance of our motion planning algorithm for randomly distributed initial positions and randomly distributed target points.  However, galaxies are not randomly distributed, but clustered.  Therefore, in a realistic situation, positioners will move toward very close targets and consequently start the next set of observation from a very close position toward other set of target points. Therefor, It is expected that adjacent positioner robots will need sets of complex maneuvers in order to find a collision-free path toward their target points.  Fig. 5 shows snap-shots of a simulation set where three positioners are engaged in a very close space. The motion planning algorithm succeeds in solving the conflict by directly executing the complex maneuvers from DNF and taking positioner robots to their target points. The main advantage of this method is that these type of conflicts could be solved in a decentralized manner which significantly decreases simulation time and motion duration. Therefore, the algorithm is reliable for large number of positioners. 

\section{Conclusion}

In the near future fibre-fed spectrograph robots such as the ones envisioned in DESI (5000 positioner robots) or PFS (24000 positioner robots) will provide a 3D map of a large portion of our universe. The main concept  which is common between the designs is the use of small mechanical robot positioners. These robot positioners are responsible for moving the fiber-ends toward their target points where they can observe different sets of objects such as galaxies, quasars or stars. As the robot positioners share work-space, the key challenge is designing a motion planning algorithm which guarantees collision avoidance. Our proposed decentralized method for coordination of positioners is a potential field that ensures collision avoidance as well as convergence of positioner robots (no blocking) to their target points. Simulation results show feasibility of using this method for mid-scale and large scale fiber-fed spectrograph robots. In-motion duration only lasts 2.5 seconds for any number of positioners. In addition, the massive spectroscopic projects could take advantage of  short simulation time to compute trajectories and the ability of interactive motion planning of the robots to target recently discovered transient sources at the last minute.

Our future research directions include the discretization of velocity profiles in order to ensure the feasibility of a real-time coordination for large number of positioner robots. In a framework where a centralized computer communicates with positioner robots, in order to minimize the communication time the motor velocities should be discretized to fit few bits. Moreover, we will explore the connection between motion planning and target assignment in order to further minimize the in-motion duration of positioner robots.

\begin{acknowledgements}
      LM, JPK and LJ acknowledge the support from the European Research Council (ERC) advanced grant "Light on the Dark" (LIDA).
\end{acknowledgements}

\bibliographystyle{plainnat}
\bibliography{main}

\end{document}